\documentclass[10pt,conference]{IEEEtran}
\IEEEoverridecommandlockouts
\usepackage{cite}
\usepackage{amsmath,amssymb,amsfonts}
\usepackage{algorithmic}
\usepackage{graphicx}
\usepackage{textcomp}
\usepackage[table]{xcolor}
\usepackage{tcolorbox}
\def\BibTeX{{\rm B\kern-.05em{\sc i\kern-.025em b}\kern-.08em
    T\kern-.1667em\lower.7ex\hbox{E}\kern-.125emX}}
\usepackage{booktabs}
\usepackage{url}
\usepackage{balance}
\usepackage{caption}
\usepackage{subcaption}

\usepackage{makecell}
\usepackage{threeparttable}
\makeatletter
\newif\if@restonecol
\renewcommand\footnoterule{%
	\kern-3\p@
	\hrule\@width\columnwidth
	\kern2.6\p@}

\usepackage{tcolorbox}

\begin{document}

\title{A Large-Scale Empirical Study on Industrial \\ Fake Apps}

\author{\IEEEauthorblockN{Chongbin Tang\IEEEauthorrefmark{1},
		Sen Chen\IEEEauthorrefmark{1},
		Lingling Fan\IEEEauthorrefmark{1},
		Lihua Xu\IEEEauthorrefmark{2},
		Yang Liu\IEEEauthorrefmark{3},
		Zhushou Tang\IEEEauthorrefmark{4},
		Liang Dou\IEEEauthorrefmark{1}\thanks{Chongbin Tang and Sen Chen are co-first authors.} \thanks{Lingling Fan and Liang Dou are the corresponding authors.}\thanks{Emails: chongbin.tang@stu.ecnu.edu.cn, ecnuchensen@gmail.com}
	}
	\IEEEauthorblockA{\IEEEauthorrefmark{1}East China Normal University, China
		\IEEEauthorrefmark{2}New York University Shanghai, China\\
		\IEEEauthorrefmark{3}Nanyang Technological University, Singapore
		\IEEEauthorrefmark{4}Pwnzen Infotech Inc., China
}}

\maketitle

\setlength{\textfloatsep}{2pt}

\begin{abstract}
		While there have been various studies towards {Android apps and their development},
		there is limited discussion of the broader class of apps that fall in the fake area.
		Fake apps and their development are distinct from official apps and belong to the mobile underground industry.
		Due to the lack of knowledge of the mobile underground industry, fake apps, their ecosystem and nature still remain in mystery.

		To fill the blank, we conduct the first systematic and comprehensive empirical study on a large-scale set of fake apps.
		Over 150,000 samples related to the top 50 popular apps are collected for extensive measurement.
		In this paper, we present discoveries from three different perspectives, namely fake sample characteristics, quantitative study on fake samples and fake authors' developing trend.
		Moreover, valuable domain knowledge, like fake apps' naming tendency and fake developers' evasive strategies, is then presented and confirmed with case studies, demonstrating a clear vision of fake apps and their ecosystem.
\end{abstract}


\section{Introduction}
With the growing attention of mobile markets, Android has accounted for 85.9\% of global market share~\cite{Gartner_report}. Over 1.5 million apps were released within 2017 alone.
Along with the booming of Android markets is the flourish of the mobile underground industry. \emph{Fake apps}, i.e., apps without official certificates, account for a major part of such underground industry.
Specifically, we consider fake apps as those who simulate the corresponding official ones or look almost the same as their official correspondences, with ultimate goal to elicit download or manifest malicious behaviors.
Early observation reveals fake apps come in two different forms.
The first category is called \texttt{imitators}, a group of apps with similar names or functionalities to their official correspondences so that users are fooled to download them.
While imitators are just similar to official apps, \texttt{imposters}~\cite{Andow2016ASO} refer to the category of apps that have exactly the same metadata with their official correspondences, for example, they may have the same names, icons, or version numbers, some of the imposters are even made by repackaging official apk files directly.

Such fake apps pose significant problems to not only the official developers' interest but also the end users' right.
For example, when users try to search an app for installation in market, multiple fake apps with similar names or icons will be retrieved at the same time.
As a result, the user experience of app searching and downloading is greatly affected by the fake apps in real world.

Even worse, as the doorsill to develop an app has been set low, the cost to develop a fake app is much lower than what it takes to develop a desktop program, providing an ideal hotbed for the underground industry to thrive on~\cite{wasserman2010software}. Moreover, the flexibility of Android app implementation~\cite{storydroid} contributes the fake apps' complexity.

Despite the ubiquity, little is known about fake apps and their ecosystem -- their common characteristics, the number of fake apps at large, their production process and speed, and their evasive strategies, etc.
Most research studies to date show greater interests on malware detection techniques~\cite{chen2016stormdroid,chen2018automated, chen2016towards, fan2016poster}. To the best of authors' knowledge, there exists no work in understanding fake apps, and their ecosystem.

Similarly, we witness the same deficiency in industry.
Most attention in analysis and threat reports focuses on malicious apps while neglecting the fake apps~\cite{McAfee_mobile_thread_report}.
On the other hand, the knowledge gained from desktop era regarding malicious or fake software are of less use, due to differences in operating system properties~\cite{yin2007panorama}.

In this paper, we focus on conducting a large-scale empirical study on fake apps and measuring the data on different dimensions.
Our goal is to systematically investigate the characteristics of fake apps from different perspectives, as well as quantitative analysis on fake apps.
Moreover, we aim to further unveil the developing trend of fake authors to help fake app detection and shine light on the fake apps' ecosystem nature for both academia and industry.

Conducting such study is not easy:
(1) To conduct a fair and representative study, it is essential to obtain appropriate subjects that are
both accessible and wildly accepted;
(2) To draw convincing conclusions, large-scale data is indispensable~\cite{fan2018large, fan2018efficiently};
(3) To provide scalable measurement, efficient analysis methodology must be taken~\cite{chen2018ausera, chen2018mobile}.

To address the aforementioned challenges, we first obtain representative objects according to an authoritative
ranking provided by online big data analysis service provider \texttt{\small Analysys}\footnote{\url{https://www.analysys.cn/}};
We then collaborate with our industrial partner, \texttt{\small Pwnzen Infotech Inc.\footnote{\url{http://pwnzen.com/}}}, one of the leading security companies in China, and collect over 150,000 data entries.
Among them, 52,638 fake samples are identified for further analyzing;
In order to work with the large-scale data, both traditional dynamic and static analysis are not feasible. Instead, we sought to common industrial practice, analyzing subjects' metadata. We identify and extract 8 metadata items to support our comprehensive measurement.

In summary, we make the following main contributions:

\begin{itemize}
	\item {\bf The first comprehensive empirical study on Android fake apps at a finer granularity}.
  	To the best of our knowledge, we are the first to provide the empirical study on fake apps.
  	We measure the fake apps from three different perspectives, allowing examining its nature at a finer granularity.

	\item {\bf A large-scale quantitative measurements on fake apps in the industry}.
	We collected more than 150,000 data entries to carry out this study to dig out valuable insights or suggestions for the industry.

  	\item {\bf An observation on the fakes of the most popular apps in the real world}.
  	We conducted our study based on the top 50 popular apps in China and their fake apps.
  	As these apps are some of the most popular app in the real world and their counterfeits, we consider our study objects to be representative enough.

  	\item {\bf Findings on fake apps' characteristics based on real cases}.
  	Discoveries and conclusions emerged from our measurement are further supported by real-world case studies.

\end{itemize}

\section{Android App Certificate}
Signature scheme is an important scheme in Android security, certificate is its essential component.
Logging the information of its owner (the developer), every certificate is unique. The functionality of certificates is two-fold:
(1) To inform the target device whom the developer of an apk file is.
(2) To provide tamper-proof to some certain degree.

Due to the uniqueness, on one hand, it's not hard to understand the certificate's first function. On the other hand, it's tamper-proof ability is implemented through a multiple-step-verification mechanism. During the signing process, the digest of the app content will be calculated. And then, a signing block, or a folder for verification will be calculated using the digest and the developer's certificate. Wherever the content is changed after signing, there will be an unmatch between the digest and the actual content, and thus such apk files are refused to be installed on the devices.

Naturally, if one fake developer modifies and repackages an existing app, in order to keep the consistency, he has to replace the original certificate with his own certificate and sign the apk file again. By checking certificates, we can easily find out the repackaged apk files (i.e. some of the \texttt{imposters}), let alone the other \texttt{imposters} and \texttt{imitators} which are totally developed by fake developers.

\begin{figure*}
	\centering
	\includegraphics[width=0.95\textwidth]{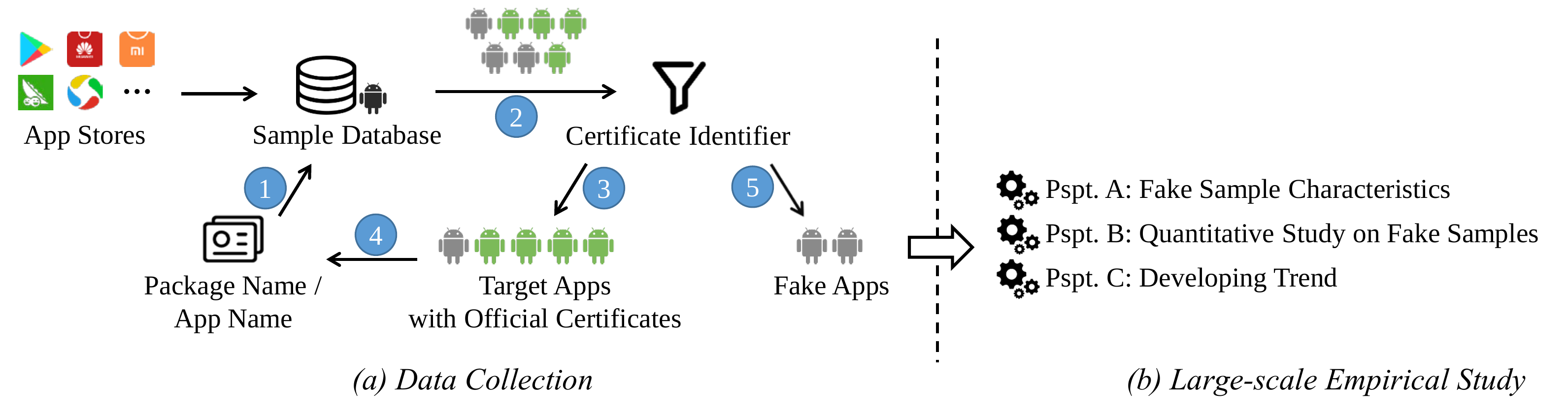}
	\caption{Workflow of our study (Pspt refers to perspective)}
	\label{fig:Workflow}
	\vspace{-3mm}
\end{figure*}

\section{Overview and Data Collection}
\subsection{Workflow of Our Study}
Fig.~\ref{fig:Workflow} shows the workflow of our study, which can be divided into two main phases: (1) \emph{Data Collection}, which collects fake apps from all sorts of app markets based on the maintained official certificates, as well as the metadata of apps (e.g., package names and app names).
Our collaboration with \texttt{\small Pwnzen} allows us to have access to raw data from most of the mainstream Android markets.
We focus on collecting the fakes of the top 50 popular apps in the real world;
(2) \emph{Large-scale Empirical Study}, which measures the collected fake apps from several perspectives: the characteristics of fake apps, as well as the quantitative analysis of fake apps.
We also aim to unveil the developing trend of fake authors to help fake app detection and shine light on the nature of the fake apps' ecosystem for both academia and industry.

\subsection{Data Collection}
Although the research community is in great need of both a comprehensive dataset on Android fake apps from industry and an effective approach to retrieving and collecting fake apps at scale from industry, little has been done to fulfill the need.
In this paper, we make the first attempt to systematically collect data from industry.

\noindent {\bf Collection Method.}
Obtaining an ample set of data is a challenging task, new app samples and updates need to be continuously crawled from various Android markets.
The challenge here is two-fold:
(1) Obtaining a large number of samples from different markets separately is no easy task;
(2) A certificate identifier is needed to tell fake apps from official ones.

To address challenge (1), we collaborate with our industry partner and leverage the Pwnzen platform\footnote{\url{https://www.appscan.io/}} to conduct \emph{sample crawling} from markets and build a \texttt{\small sample database};
to meet challenge (2), we pre-download the latest samples of our target apps to extract their official certificate information and construct a certificate identifier.

More specifically, during the database construction phase, we clusters samples by certificate hash, package names or app names.
When the database receives queries, it returns sample clusters with the corresponding package names, app names, certificate hash, etc, in form of sample metadata entries.
Since fake samples usually have similar names to the official ones, we collect fake apps by using app names from official apps.
To achieve this, we first extract the package names of our pre-downloaded samples.
These package names are later sent to the sample database for query (i.e., step 1 marked in Fig.~\ref{fig:Workflow}).
And then, for each metadata entry returned from step 2, we check if it is official using the certificate identifier.
If true and that sample is confirmed as one of our target app (step 3 in the figure), namely, it has the same package name to one of the inputs, its name will be recorded (step 4) and used for further query (back to step 1).
If false, that sample would be marked as a fake one (step 5), its metadata will be utilized for large-scale measurement.
For each app, once all of the official names have been used for query, the data collection on it is finished.

\noindent {\bf Collected Dataset.}
Here is a bird eye's view to the data we collected:
we chose the top 50 popular apps from Analysys's ranking, within 11 app categories, as our target apps. Since apps may change their names over time,  we recorded 198 app names from the 50 apps to mine fake samples.
Among the 50 apps, we failed to find any samples from the following three apps: \emph{OPPO AppStore}, \emph{Huawei AppStore}, and \emph{MI AppStore}, because they are developed by cellphone manufacturers and are not provided to other app markets.
This is also the reason why three apps are popular -- they are preinstalled into every single device produced by their manufacturers.
Thus we finally obtained 47 target apps in total.
With the 47 target apps, we retrieve 138,106 distinct samples in total, 69,614 of which are official samples of our target apps, 52,638 samples lack registered certificates.
For each sample, we retrieve 8 data items as metadata: \emph{Sample SHA1}, \emph{Certificate SHA1}, \emph{Package Name}, \emph{Package Size}, \emph{Version Number}, \emph{Retrieved Time}, and \emph{Source}. Among them, \emph{Sample SHA1} and \emph{Certificate SHA1} are the hash code for APK files and certificates under SHA1 algorithm respectively. \emph{Retrieve Time} tells when the sample was crawled from app store and \emph{Source} tells which store the sample is from.

Empirical study is then applied to these metadata, especially to those of fake apps, to gain us a more comprehensive understanding on fake apps' nature and characteristics, and the behaviors of fake app authors.

\section{Large-scale Empirical Study and Discoveries}

With the large-scale dataset ready, we can further conduct a comprehensive empirical study to acquire the nature of fake apps as well as understanding their ecosystem.
To effectively measure different facades of fake apps, We define three perspectives, namely \emph{Fake Sample Characteristics}, \emph{Quantitative Study on Fake Samples}, and \emph{Developing Trend}.
Next, we'll describe each perspective in detail.

\subsection{Fake Sample Characteristics}
To reveal the strategy the fake app authors are employing, or how they bypass app markets' security scheme, fake sample characteristics have to be understood.
As such, we conduct our measurement in terms of certificates and basic information like app names, package names and package sizes.

Certificate serves as the identifier for developers.
The nature of the certificate, namely, whether each fake app has a unique certificate, is likely to be essential to fake apps' evasive technique.
On the other hand, we believe repackaged apps, as a kind of \texttt{imposters}, are widespread in our dataset.
Measurement on basic information of fake apps, such as package names and package sizes, helps us determine how repackaged apps are distributed, since repackaging an app does not change any of its basic information (i.e. the app name, package name, version code, etc.) unless it's done intentionally.

To this end, we have some hypotheses as below:

\noindent{\bf Hypo 1.1:} Most of these fake samples have their corresponding unique certificates.
In other words, most fake certificates and fake samples have a one-to-one relation.

\noindent{\bf Hypo 1.2:} A large portion of fake samples have the same app names/package names/apk sizes as those in official samples.

To verify our hypotheses as well as to gain knowledge to fake sample individuals, we propose the following research questions in this subsection.

\noindent{\bf RQ 1.1}: What's the relationship between the number of fake samples and their certificates? That is, how many fake samples does one certificate usually link to?

\noindent{\bf RQ 1.2}: How do fake apps imitate official apps? That is, how similar are the names/package names/apk sizes of fake samples compared to those of official samples?

\begin{figure*}
	\centering
	\begin{subfigure}[t]{0.33\textwidth}
		\centering
		\includegraphics[width=1\textwidth]{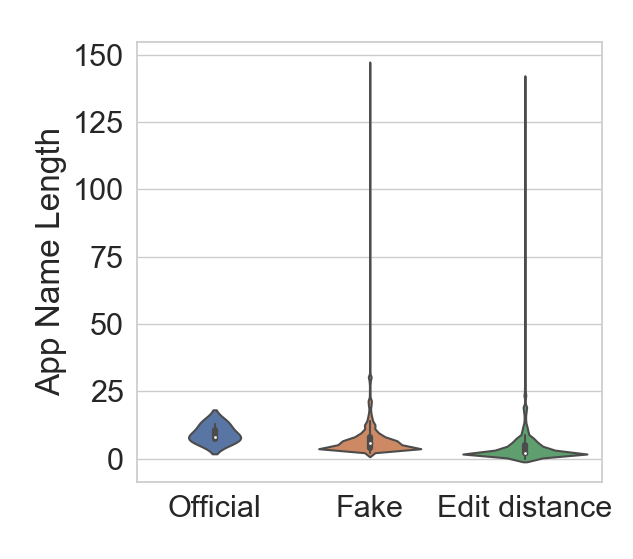}
		\caption{App name}
		\label{fig:appname}
	\end{subfigure}%
	\begin{subfigure}[t]{0.33\textwidth}
		\centering
		\includegraphics[width=1\textwidth]{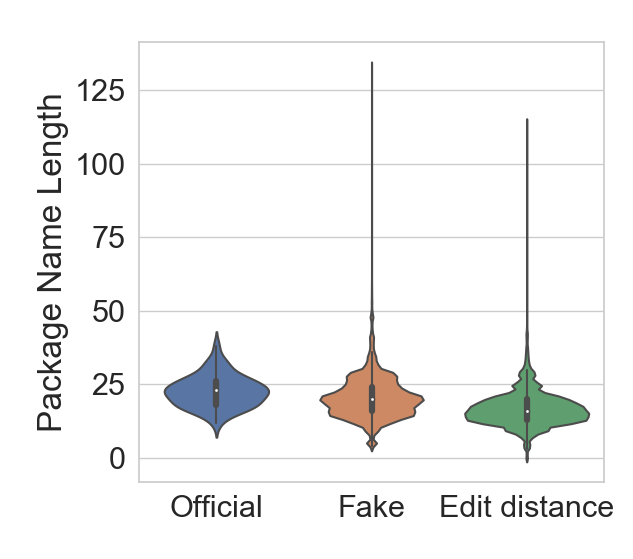}
		\caption{Package name}
		\label{fig:pkgname}
	\end{subfigure}%
	\begin{subfigure}[t]{0.33\textwidth}
		\centering
		\includegraphics[width=1\textwidth]{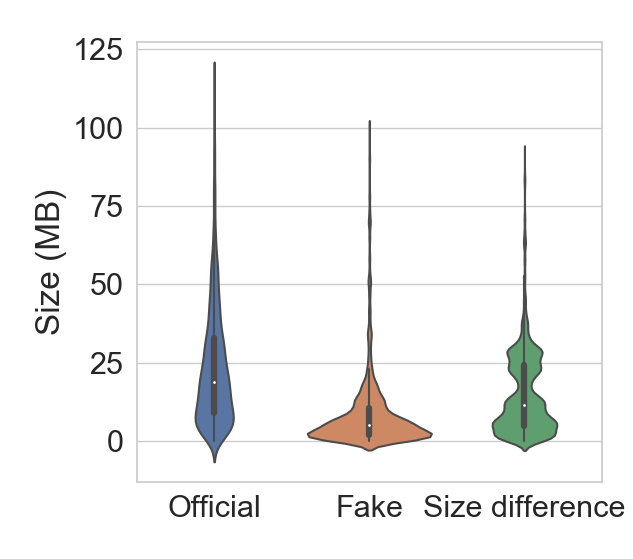}
		\caption{Size}
		\label{fig:size}
	\end{subfigure}%

	\caption{Statistic on app attributes}
	\label{fig:Statistic_fake_and_official}
	\vspace{-5mm}

\end{figure*}

\noindent{\bf Answer to RQ 1.1.}
76\% of these fake certificates are linked to merely one or two fake samples, and the number of fake examples a certificate links to is various from 1 to 1,374.
We count the number of certificates which link to different sample number in table~\ref{table:certificate_number_statistic}.

\begin{table}
  \renewcommand{\arraystretch}{1}
  \footnotesize
  \centering
  \caption{Statistics on fake samples and their certificates}
  \vspace{1mm}
  \begin{tabular}{l c c c c c c c}
  \toprule
  {\bf \# of samples} & {\bf 1-5} & {\bf 6-10} & {\bf 11-50} & {\bf 51-100} & {\bf More than 100} \\
  \midrule
  {\bf \# of certificates} & 8252 & 525 & 531 & 71 & 80 \\
  \bottomrule
  \end{tabular}
  \label{table:certificate_number_statistic}
\end{table}

This discovery partly matches our assumption that most of these fake samples have their corresponding unique certificates.
We consider this as a strategy to bypass app markets' security scheme, as even if one fake sample is exposed, other fake samples developed by the same developer will not be implicated directly.
Nevertheless, when reviewing certificates linked with multiple fake samples, we find some very surprising findings that we will expound in Section~\ref{sec:casestudy}.

\noindent{\bf Answer to RQ 1.2.}
According to our statistical result, only 243 out of 52,638 samples (less than 0.5\%) use official package names, all the rest fake samples (more than 99.5\%) use their own package names.
In the rest 52,395 samples, 14,089 different package names were found.
But does this mean fake samples are all using package names that are totally different from the official ones? Could they be using package names that are similar to their official correspondences?

To figure out the similarity, we utilize \textit{edit distance}~\cite{levenshtein1966binary}, a distance definition widely applied in natural language processing (NLP):
{Given two strings $a$ and $b$, the edit distance $d(a, b)$ is the minimum-weight series of edit operations that transform $a$ into $b$.
In our case, edit operations refer to either to append, to delete or to change a character.}
For instance, the edit distance between string ``fake" and ``official" is 7, while between ``jingdong" and ``jindeng", this value becomes 2.
For every fake package name from a fake sample, we compute its edit distance to the official package name of its original.

Fig.~\ref{fig:Statistic_fake_and_official} is consist of three violin plots,\footnote{\url{https://en.wikipedia.org/wiki/Violin_plot/}} representing our statistics on app names, package names and package sizes, respectively.
In each ``violin'', the white dot represents the median, the thick bar in the middle represents the interquartile range while the thin bar represents 95\% confidence interval.

Fig.~\ref{fig:appname} shows the statistic information on app names of official samples, fake samples, and the edit distance between them.
Both the white dot in ``Official'' violin and the one in ``Fake'' violin are at a similar level near the value ``6'', which means the average length of app names of both official samples and fake samples are close to each other.
The overall distribution of these two data groups have similar bodies, signals that they are also similar as well.
What's more, the median value of edit distance is low (``2'' on $y$-axes), meaning that half of the fake apps get their names by modifying less than 3 characters from the corresponding official apps' names.
This is a proof indicating that most fake apps are using a similar name to an official name.
At the same time, we notice that some fake apps have pretty long names (there is one with a name of 146-character-long length).
Many of those outliers are samples uploaded by fake authors, maybe for testing purpose to explore the vetting mechanism. The other purpose is to associate users' search keywords as far as possible.

Fig.~\ref{fig:pkgname} shows the result on package names.
Like the plots in Fig.~\ref{fig:appname}, the difference between the average length of package names of official apps and the average length of package names of fake apps is still tiny (they are of value ``23'' and ``20'', respectively).
Nonetheless, the median of edit distance between them is explicitly higher (``16'' on $y$-axes), which means it takes averagely 16 times modification to turn a fake package name to an official package name and vice versa.
Thus, we infer that fake apps tend to use self-defined package names.

Fig.~\ref{fig:size} reports package size information.
To better represent the trend, we eliminated some outliers: samples that are larger than 150MB (851 in 69,614 official samples (about 1\%) and 447 in 52,638 fake samples (less than 1\%), most of which are from game category).
The figure shows that the median number of fake samples' size is around 5MB, while half of the official apps have a size greater than 18MB, meaning that fake apps are more likely to be
(1) developed by their owners but not originated from repackaging official apps,
(2) malicious apps, for malicious apps are usually in small sizes.

In short, Fig.~\ref{fig:Statistic_fake_and_official} tells that fake apps
(1) prefer to use a similar (or even same) name to an official app's name, but they have their own package names and
(2) are usually of a small size.

To a large extent, we owe the first point to the incompleteness of the information the app store displays on apps.
In most app stores, when users browse an app's detail page, they can only see the app's name, description, user comments and ratings which are positive for leading users to download that app.
However, technical information rarely appears.
In some app markets, users don't even know how large an apk file is.

\vspace{1mm}
\noindent\fbox{
	\parbox{0.95\linewidth}{
		\textbf{Remark 1}: Most certificates link with only a number of fake apps, which is highly possible to be a fake developers' evasive strategy.
			Moreover, we observe that fake apps do tend to use official app names or names alike.
			Nonetheless, fake apps and official apps are not resemble in terms of package names or apk sizes, disclosing that repackaged apps are not mainstream in fake apps.
		
	}
}

\subsection{Quantitative Study on Fake Samples}
It is valid to assume that fake app developers are driven by profits, hence there is a likelihood that the number of fake app is correlated to their source market, popularity and categories.
In addition, the update frequency can be taken in as a factor, too.
Accordingly, we hypothesize the following factors may influence the number of fake samples of an app:

\noindent{\bf Hypo 2.1:} {The rate of fake samples is related to the number of apps a market contains.}

\noindent{\bf Hypo 2.2:} The number of fake apps are closely related to how popular an app is.

\noindent{\bf Hypo 2.3:} Update frequency effects the number of fake samples.

\noindent{\bf Hypo 2.4:} Category is a factor influencing the fake sample number.

Correspondingly, we define our research questions as follows:

\noindent{\bf RQ 2.1}: Where are these fake samples mainly from?

\noindent{\bf RQ 2.2}: Does the popularity of an app affect the number of its fake samples?

\noindent{\bf RQ 2.3}: Does an app's update frequency influence its fake sample's number?

\noindent{\bf RQ 2.4}: Is the number of fake samples related to the app's category?

\begin{table*}
\renewcommand{\arraystretch}{1.05}
\footnotesize
\centering
\setlength{\belowcaptionskip}{-5pt}
\caption{Our target app and their related statistics}
\vspace{1mm}
\rowcolors{2}{gray!25}{white}
\label{table:data-statistics}
\begin{threeparttable}
\begin{tabular}{l l c c c c c c}
\toprule
{\bf Name} & {\bf Category} & \begin{tabular}[c]{@{}c@{}}{\bf MAI} {\bf (Monthly} \\ {\bf Activeness Indicator)} \end{tabular} & \begin{tabular}[c]{@{}c@{}}{\bf Update Frequency} \\ {\bf (day/version)} \end{tabular} & {\bf \#Total} & {\bf \#Fake} & \begin{tabular}[c]{@{}c@{}}{\bf Fake Sample} \\ {\bf Rate} \end{tabular} & \begin{tabular}[c]{@{}c@{}}{\bf Avg Fake Latency} \\ {\bf (day)} \end{tabular} \\
\midrule
{\bf WeChat}\tnote{*} & {\bf SocialNetwork} & {\bf 91.2K} & {\bf 6.4} & {\bf 9248} & {\bf 6447} & {\bf 69.7\%} & {\bf 12.1} \\
{\bf QQ}\tnote{*} & {\bf SocialNetwork} & {\bf 54.6K} & {\bf 10.7} & {\bf 11167} & {\bf 3780} & {\bf 33.8\%} & {\bf 9.2} \\
iQiyi & Video & 53.5K & 6.4 & 7586 & 3481 & 45.9\% & 9.3 \\
Alipay & Life & 48.1K & 10.2 & 983 & 231 & 23.5\% & 10.1 \\
{\bf Taobao}\tnote{*} & {\bf OnlineShopping} & {\bf 47.5K} & {\bf 7.0} & {\bf 6003} & {\bf 3010} & {\bf 50.1\%} & {\bf 8.1} \\
TencentVideo & Video & 47.3K & 6.3 & 1429 & 68 & 4.8\% & 10.7 \\
Youku & Video & 40.9K & 7.3 & 2058 & 262 & 12.7\% & 6.7 \\
{\bf Weibo}\tnote{*} & {\bf SocialNetwork} & {\bf 39.2K} & {\bf 5.3} & {\bf 5947} & {\bf 2715} & {\bf 45.7\%} & {\bf 5.7} \\
WiFiMasterKey & SystemTool & 36.4K & 3.1 & 4808 & 2999 & 62.4\% & 3.0 \\
SougouInput & SystemTool & 33.3K & 11.0 & 898 & 40 & 4.5\% & 21.8 \\
MobileBaidu & Information & 32.4K & 11.1 & 15651 & 3514 & 22.5\% & 12.8 \\
TencentNews & Information & 28.7K & 8.5 & 1051 & 11 & 1.0\% & 8.9 \\
QQBrowser & Information & 27.8K & 5.6 & 1369 & 43 & 3.1\% & 11.6 \\
Toutiao & Information & 27.4K & 4.4 & 3538 & 179 & 5.1\% & 5.6 \\
Myapp & AppStore & 27K & 11.4 & 2419 & 266 & 11.0\% & 11.6 \\
Kwai & Video & 24.4K & 3.2 & 8273 & 4270 & 51.6\% & 3.5 \\
WeSecure & SystemTool & 24.2K & 8.7 & 2463 & 1340 & 54.4\% & 8.7 \\
Amap & Life & 24K & 6.5 & 1225 & 51 & 4.2\% & 13.1 \\
KugouMusic & Music & 23K & 8.6 & 1313 & 122 & 9.3\% & 12.2 \\
QQMusic & Music & 21.7K & 9.4 & 1132 & 65 & 5.7\% & 14.6 \\
BaiduMap & Life & 21.3K & 8.8 & 2609 & 1438 & 55.1\% & 15.3 \\
TikTok & Video & 19.4K & 11.1 & 317 & 12 & 3.8\% & 8.3 \\
{\bf JD}\tnote{*} & {\bf OnlineShopping} & {\bf 18.5K} & {\bf 10.9} & {\bf 5000} & {\bf 2377} & {\bf 47.5\%} & {\bf 12.3} \\
UCBrowser & Information & 16.7K & 7.4 & 4232 & 1624 & 38.4\% & 7.0 \\
360Security & SystemTool & 15.4K & 12.4 & 3670 & 1423 & 38.8\% & 19.1 \\
TencentKaraoke & Music & 14.7K & 21.1 & 618 & 215 & 34.8\% & 17.3 \\
MeiTuan & Life & 13K & 8.0 & 4752 & 1415 & 29.8\% & 6.9 \\
{\bf Pinduoduo}\tnote{*} & {\bf OnlineShopping} & {\bf 12.9K} & {\bf 6.6} & {\bf 2327} & {\bf 551} & {\bf 23.7\%} & {\bf 7.8} \\
{\bf ArenaofValor}\tnote{*} & {\bf Game} & {\bf 12.5K} & {\bf 15.5} & {\bf 2350} & {\bf 1319} & {\bf 56.1\%} & {\bf 12.3} \\
MeiTuXiuXiu & Camera & 12.4K & 5.4 & 1705 & 784 & 46.0\% & 5.8 \\
VigoVideo & Video & 12.2K & 11.9 & 410 & 16 & 3.9\% & 9.6 \\
MojiWeather & Life & 12K & 4.2 & 10081 & 7093 & 70.4\% & 4.7 \\
DiDi & Life & 11.8K & 8.6 & 943 & 117 & 12.4\% & 7.0 \\
HuaweiAppStore & AppStore & 11.8K & N/A & 0 & 0 & 0.0\% & N/A\\
{\bf HappyElements}\tnote{*} & {\bf Game} & {\bf 11.2K} & {\bf 19.7} & {\bf 2406} & {\bf 1738} & {\bf 72.2\%} & {\bf 20.6} \\
KuwoMusicPlayer & Music & 11K & 2.9 & 3778 & 69 & 1.8\% & 4.2 \\
iXigua & Video & 11K & 11.5 & 866 & 100 & 11.5\% & 8.8 \\
OPPOAppStore & AppStore & 10.8K & N/A & 0 & 0 & 0.0\% & N/A\\
CleanMaster & SystemTool & 9.9K & 10.3 & 1803 & 388 & 21.5\% & 13.5 \\
360CleanDroid & SystemTool & 9.6K & 17.3 & 327 & 8 & 2.4\% & 8.5 \\
360Zhushou & AppStore & 9.2K & 7.6 & 1616 & 137 & 8.5\% & 8.4 \\
TencentWiFiManager & SystemTool & 8.8K & 19.5 & 1636 & 658 & 40.2\% & 15.7 \\
XunfeiInput & SystemTool & 8.6K & 6.0 & 1451 & 8 & 0.6\% & 10.1 \\
BaiduAppSearch & AppStore & 8.2K & 11.4 & 3849 & 437 & 11.4\% & 14.5 \\
MiAppStore & AppStore & 7.8K & N/A & 0 & 0 & 0.0\% & N/A\\
{\bf WPSOffice}\tnote{*} & {\bf Productivity} & {\bf 7.4K} & {\bf 6.0} & {\bf 1152} & {\bf 69} & {\bf 6.0\%} & {\bf 7.8} \\
BeautyCam & Camera & 7.1K & 5.3 & 1600 & 691 & 43.2\% & 6.3 \\
NeteaseCloudMusic & Music & 7K & 10.5 & 616 & 6 & 1.0\% & 12.2 \\
NeteaseNews & Information & 6.7K & 7.0 & 1441 & 93 & 6.5\% & 5.0 \\
{\bf QQMail}\tnote{*} & {\bf Productivity} & {\bf 6.6K} & {\bf 16.4} & {\bf 520} & {\bf 11} & {\bf 2.1\%} & {\bf 10.4} \\
\bottomrule
\end{tabular}
\begin{tablenotes}
  \footnotesize
  \item[*] Detailed descriptions are given in {\bf Answer to RQ 2.4}
\end{tablenotes}
\end{threeparttable}
\vspace{-3mm}
\end{table*}

\begin{figure*}
	\centering
  \setlength{\belowcaptionskip}{-10pt}
	\includegraphics[width=\textwidth]{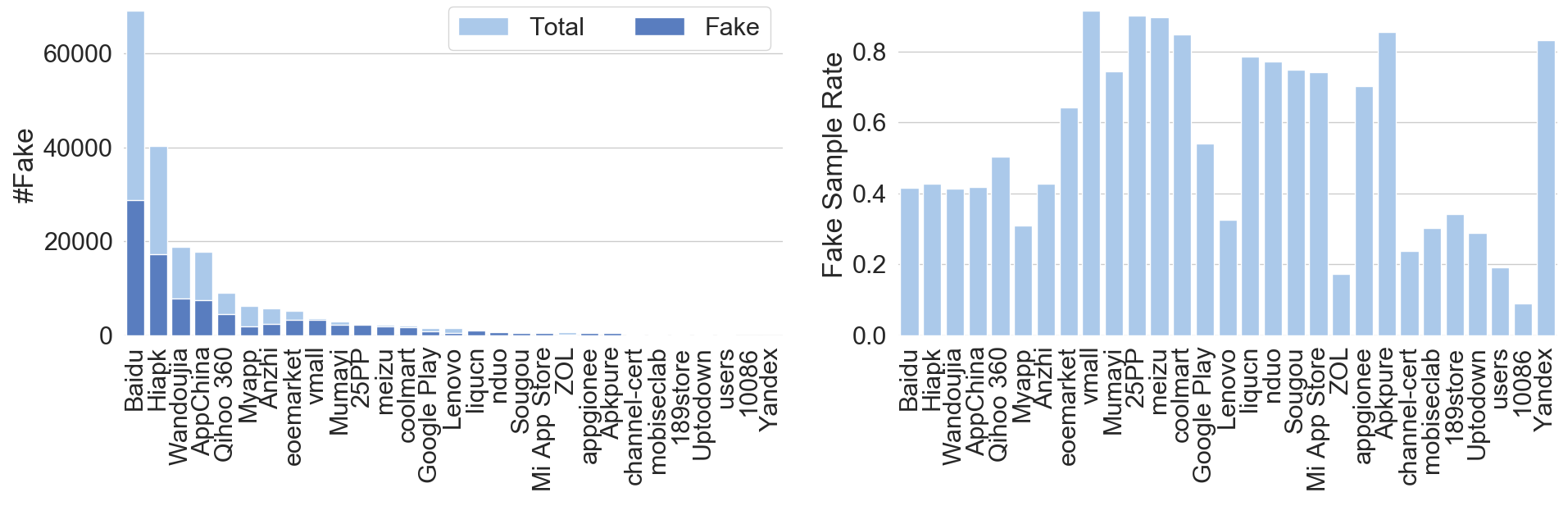}
	\caption{Number of samples collected from different markets}
	\label{fig:Sample_source}
\end{figure*}

\noindent{\bf Answer to RQ 2.1.} Fig.~\ref{fig:Sample_source} shows the samples' origin.
From the left subplot, \texttt{Baidu App Store} not only provides the largest sample number among all 31 different app sources, but is also the source where most fake samples are from.
Fake sample rates are displayed on the right subplot.
Although both \texttt{Baidu App Store}~\cite{Baiduappstore} and \texttt{Hiapk}~\cite{Hiapk} hold a fake sample rate of about 40\%, the number of fake samples from \texttt{Baidu App Store} exceeds \texttt{Hiapk} to a great extent due to its dominant total sample number.
Although no connection between the number of fake samples and market can be found from our data, we notice that the relationship between apps and markets may affect the fake rate.
This is well supported by the low fake rate of \texttt{Myapp}~\cite{Myapp} -- the app market provided by \texttt{Tencent}, which is also the 12 out of 50 developers in our target apps.

\noindent{\bf Answer to RQ 2.2.}
Intuitively, the more popular an app is, the more possible it would get shammed, for fake developers would mislead users to download their apps to gain profits.

Note that each app has different amount of samples (including official samples and fake samples), processing our measurement directly based on the number of fake samples is incorrect.
To counteract this bias, each fake count should be regularize into a \textit{fake sample rate}, the rate of fake samples in all collected samples of an app.

Next, we employ a metric called \textit{Pearson product-moment correlation coefficient (PPMCC)} to reveal relativeness between an app's fake sample number and its popularity, which uses the regularized fake sample rates and monthly activeness indicators (MAI) obtained from Analysys~\cite{yiguanqianfan}.
This value ranges from -1 to 1, the closer the PPMCC value is to 0, the weaker correlation between the two factors is indicated.
Surprisingly, according to our data, the value of PPMCC between this two factors is 0.246, revealing that the fake sample number and an app's popularity only hold their relativeness on a weak level, which does not match our expectation.

\noindent{\bf Answer to RQ 2.3.}
We assume the update frequency is related to the number of fake samples of an app, for updates can usually help keep a software from being attack.
The higher the update frequency is, the safer an app is supposed to be.

To estimate the average update frequency of our target apps, the time when an app's official sample was crawled and when its latest official samples were crawled is marked.
The difference between them is then divided by the number of that app's existing version to obtain an update frequency, with unit day/version.

The result PPMCC value of 0.084 shows that the connection between an app's update frequency and its fake sample rate barely exists.
We attribute this result to two reasons:
(1) The high update frequency (10 days/version on average for apps in our dataset) indicates app developers may not fix security issues in per update, weakening the function of update frequency as a security indicator.
(2) A large portion of fake samples in our dataset are not derived from repackaging. To this end, fake developers can produce fakes regardless of how well the official apps are protected.

\noindent{\bf Answer to RQ 2.4.}
Some categories are potentially more profitable than others.
A report from the app marketing company LIFTOFF~\cite{LIFTOFF_report} forecasts gaming to be the next most billable area.

Our 50 target apps are divided into 11 categories according to their functionalities, Table~\ref{table:data-statistics} shows these categories and their corresponding fake sample rate.
In the same category, the difference between apps on fake rate lies in an acceptable range.
Without doubt, entertainment related categories like \texttt{Game} and \texttt{Social Network} attract more fake samples.
Another field, \texttt{Online shopping}, has also gained special love from fake developers because online shopping is rapidly developing in China.
Relatively, \texttt{Productivity} is not that attractive to fake developers, the average fake sample rate of this filed is only 4.05\%.
Apps in these four categories are marked in bold in the table.

The result matches the observation in our daily life, people always tend to use mobile devices for entertainment instead of business purpose.
It's pretty interesting to discover that the number of fake samples in a way reflects how people use their phone in their daily lives.

\vspace{1mm}
\noindent\fbox{
	\parbox{0.95\linewidth}{
		\textbf{Remark 2}: As revealed by statistics, the number of samples returned from an app store does not imply a fake rate.
		Additionally, the relationship between apps and market itself influences the number of fake samples from that market.
		To our surprise, an app's update frequency is not tightly correlated with its fake rate.
		We owe this to the fact that apps are updated too frequently and that repackaged samples are of minority in our dataset.
		We further observe that ``category'' as a factor has greater influence on the number of fake samples of an app than ``popularity'' and ``update frequency''. 
		
	}
}

\begin{figure}
	\centering
	\includegraphics[width=0.5\textwidth]{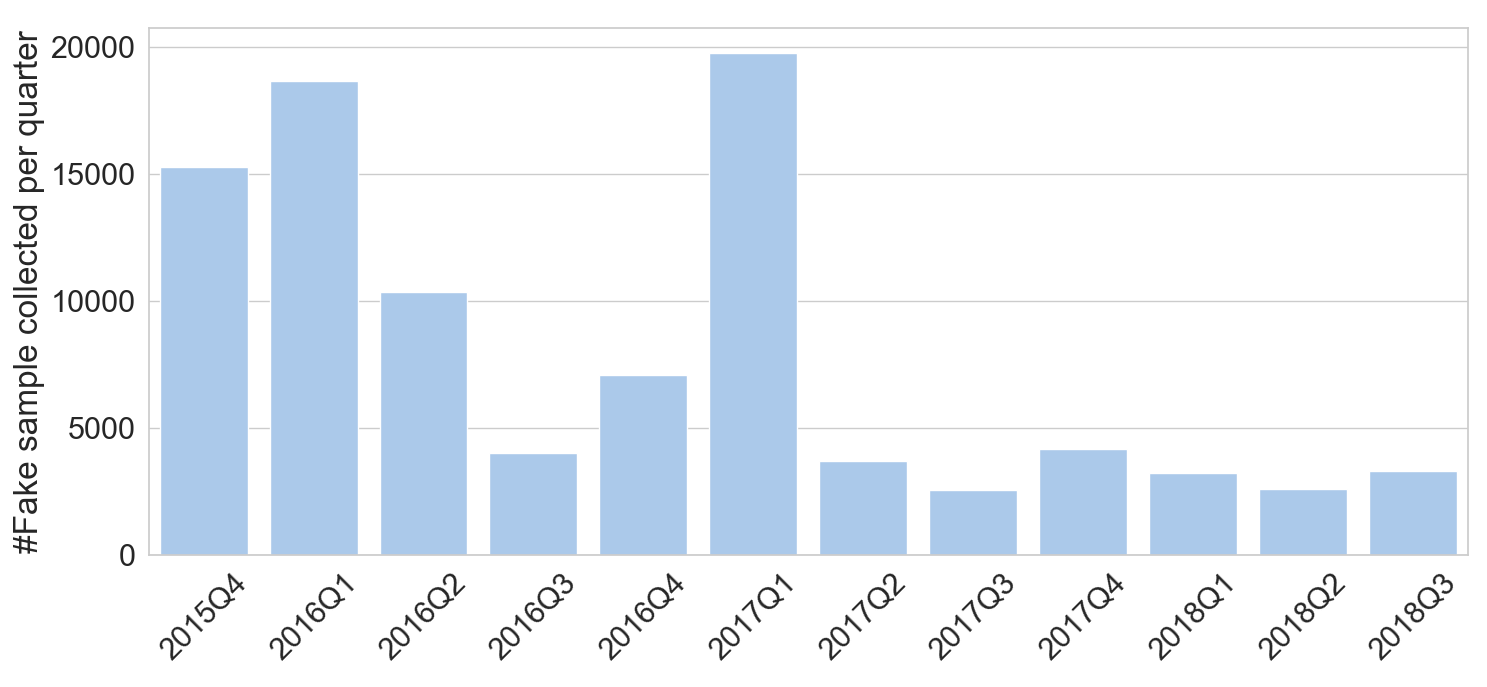}
	\caption{Numer of fake samples collected per quarter}
	\label{fig:Number_per_quarter}
\end{figure}

\subsection{Developing Trend}
In order to figure out fake apps' characteristics or behavior patterns over time, we propose the following research questions:

\noindent{\bf RQ 3.1} After a new version of an official app is published, how long do fake developers take to publish a new fake sample? In other words, how soon will these copycats appear?

\noindent{\bf RQ 3.2} How long can a fake app's certificate survive?

\noindent{\bf RQ 3.3} Is there a changing pattern of fake samples over time?

\begin{figure}
	\centering
	\includegraphics[width=0.5\textwidth]{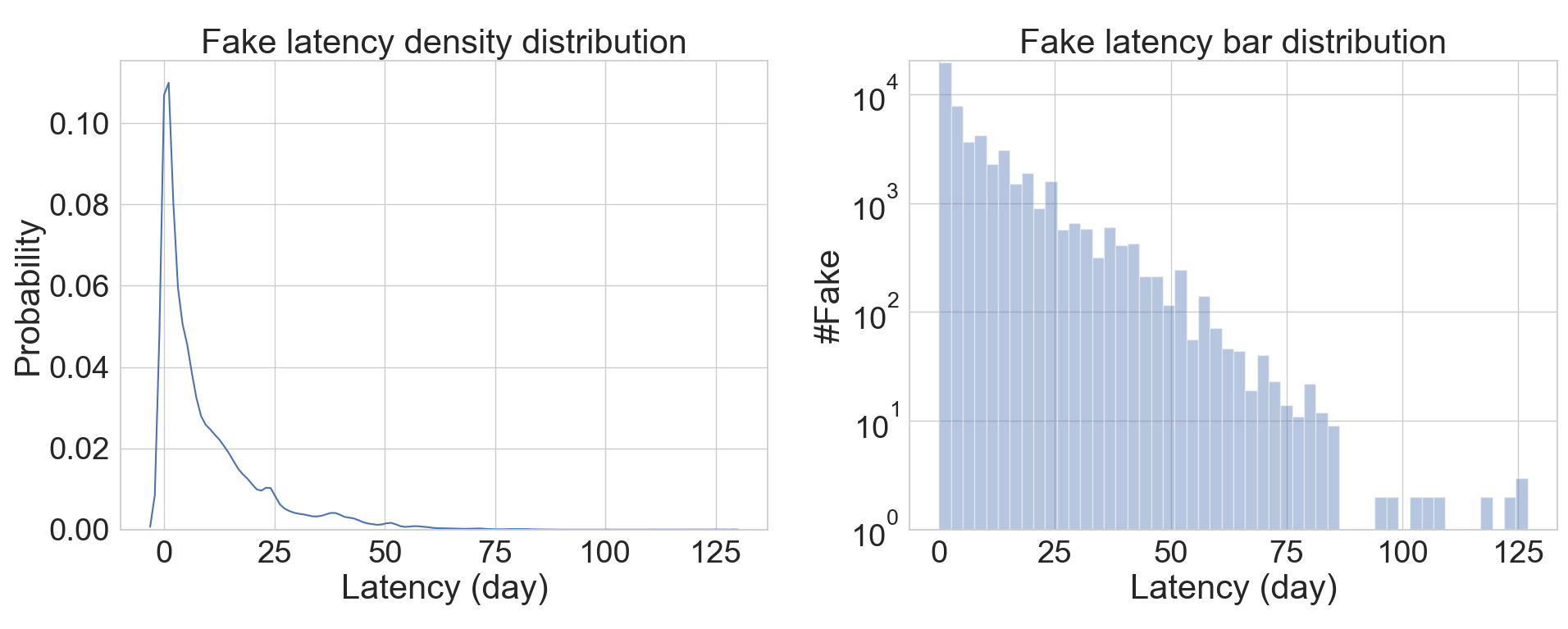}
	\caption{Fake latency overall distribution}
	\label{fig:Fake_latency_overall_distribution}
\end{figure}

\noindent{\bf Answer to RQ 3.1.}
We compute this latency and show its distribution in Fig.~\ref{fig:Fake_latency_overall_distribution}.

Due to various reasons, it is hardly possible to retrieve the complete updating timeline for every single official app in our study, yet we approximately reproduce them with our data.
Firstly, we categorized all the official samples by their origins, and further categorized samples in each origin by version number.
After that, for each app and each version the samples are sorted by the date they were crawled, so by extracting the crawled date of the first sample in each version, we can obtain the earliest date a version is released.
Lastly, by combining and sorting the release dates of different versions according to different apps, we can reproduce the updating timelines of our target apps.

To find out the release latency of a fake app, all the dates on the timeline of the corresponding official app are compared in order to find out the smallest negative difference which we define as the release latency.
Fig.~\ref{fig:Fake_latency_overall_distribution} shows that most fake samples are published with the latency shorter than 20 days.
According to our statistics, 60\% of fake samples show up in 6 days after a new version of the official app is published.
This reveals a truth that fake developers are swift in action.

\begin{figure}
	\centering
	\includegraphics[width=0.5\textwidth]{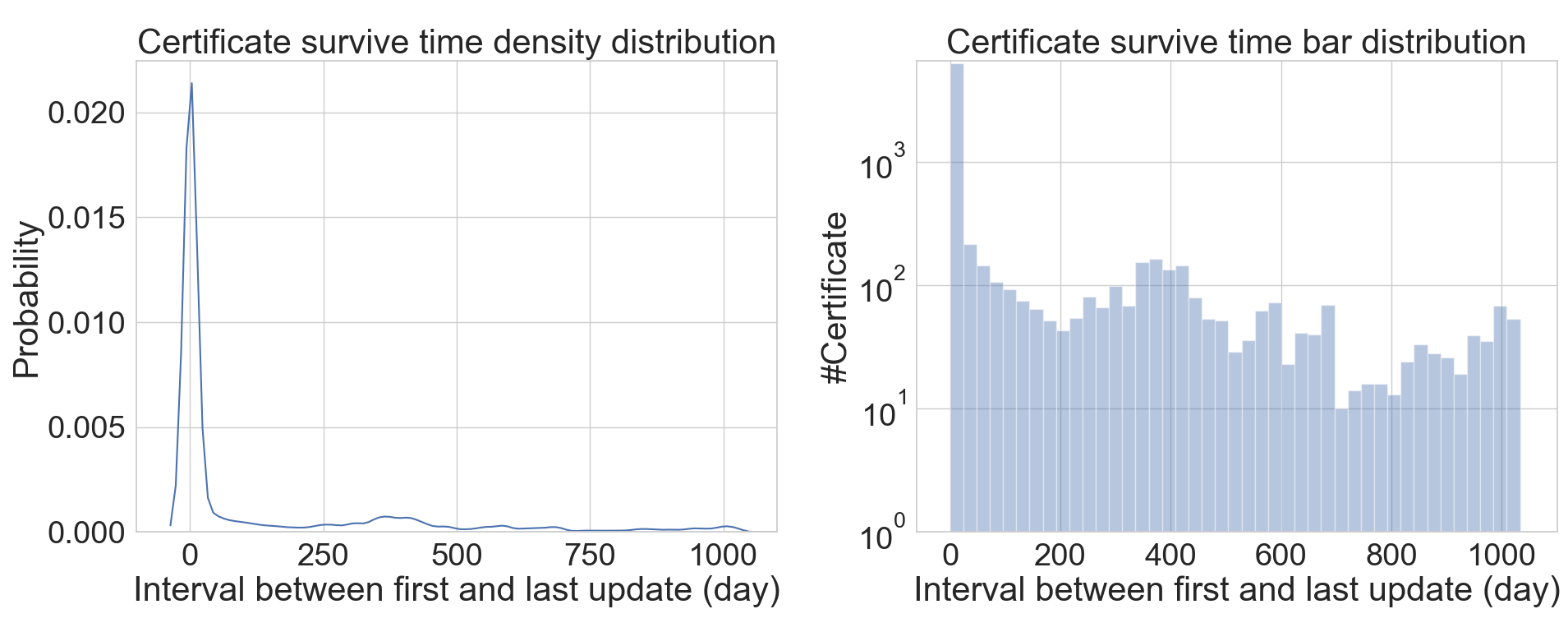}
	\caption{Fake certificate survival time distribution}
	\label{fig:Fake_certificate_survival_distribution}
\end{figure}

\noindent{\bf Answer to RQ 3.2.} Fig.~\ref{fig:Fake_certificate_survival_distribution} shows the distribution of the time a fake certificate can survive in markets.
In the left density distribution subplot, $x$-axes is the latency and $y$-axes shows the probability density of data at corresponding $x$ value.
The total area under the curve is 1, and the area under two $y$ values $y1$ and $y2$ is the probability of their corresponding value $x1$ and $x2$ account for in data.
For example, in Fig.~\ref{fig:Fake_certificate_survival_distribution}, the area beneath curve between 0 to 200 on $x$-axes is close to 0.8, which means nearly 80\% of certificates only survive for no more than 200 days.

To judge how long a fake certificate can survive is similar to how we calculate the update frequency of an app, the first time and the last time a fake sample from the same certificate gets crawled are marked.
The time when a sample was crawled from a market might be different from the time when it is available in the market, but our crawler downloads new samples from different markets by days and we also use days as the unit in our measurement, so we can approximately regard this two values as the same one.

As shown in Fig.~\ref{fig:Fake_certificate_survival_distribution}, the distribution of fake certificate survival time shows that almost all the fake certificates live a short life, which means most fake certificates only show up in a short period of time.
This can be explained by a scheme that most markets have.
Once an app is found malicious or illegal, the market would stop that specific developer from uploading more samples by refusing to receive samples with the same certificate.
There are also a number of certificates which can survive for a long time.
According to the figure, some fake certificates even traverse the whole study interval.
We will conduct a case study on this phenomenon in Section ~\ref{sec:casestudy}.

\noindent{\bf Answer to RQ 3.3.} Fig.~\ref{fig:Number_per_quarter} shows the number of fake samples collected per quarter since the fourth quarter of 2015.
Although a large number of new fake samples get released in every quarter, the figure shows a tendency that the total number of fake apps on markets is gradually decreasing by years.
Note that our statistics only focus on fake samples, consequently this phenomenon does not indicate the underground industry is turning down.
Instead, we suppose this is possibly caused by the reform of fake apps.

On one hand, as stricter review schemes and stronger protection systems are applied on app stores, it's inevitable that fake apps in this study, become harder and harder to get on the shelf.
On the other hand, the new generation of malicious software, such as ransomware~\cite{ransomware} is impacting the underground industry.
Compare to fake apps, the new malicious apps are not only hard to defend (due to the innovative or even state-of-the-art techniques they utilize) but also extremely profitable.
Wannacry, a ransomware which was first spotted in the 2nd quarter of 2017, conquered tens of thousands of devices in a couple of weeks, which directly pulled up Bitcoin's price like a rocket~\cite{wannacry_bitcoin_news}.
Afterward, in the first quarter of 2018, a burst of cryptomining malware on phones emerged~\cite{comodo_report}.
This may be the reason why the number of fake samples suffers two suddenly drops in the second quarter of 2017 and the first quarter of 2018, respectively.

\vspace{1mm}
\noindent\fbox{
	\parbox{0.95\linewidth}{
		\textbf{Remark 3}: Fake apps can be produced in a relatively short time, and the dropping number of fake samples by years suggests that they are mired in recession.
		Besides, only a few fake certificates survive for a long time, confirming that markets' protection schemes do work to some extent. 
	}
}

\section{Case Study and Discussion}
\label{sec:casestudy}
In this section, we present some samples in our dataset, not only to firm our findings but also to provide more valuable insights.

\noindent{\bf Case study 1.} \emph{Fake certificate with multiple malicious imitators and imposters}

We manually review the samples signed by the certificate with SHA1 ``\emph{61ed377e85d386a8dfee6b864bd85b0bfaa5af81}", the certificate with the most number of fake samples among our fake certificate set (i.e., 1,374 fake samples).
On top of that, this certificate is also one of the certificates survive the longest time (nearly 3 years) and is still active.

Originally, we presume this certificate to belong to a benign app which passes the verification of analysts in Pwnzen, since the number of samples it links to even exceeds the number of official samples of some apps.
The truth is, however, after manually review, we found all the 1,374 samples linked with this certificate are typical fake samples, in form of either imitators or imposters, covering 79\% (37 out of 47) of our target apps.
Some of its samples can even be organized in version order, which means the developer does track official apps to update its fake versions as maintenance.

We display some of the samples singed by this certificate in Table~\ref{table:certificate_case_study}, they are all reported to be malicious (i.e., Ad-ware, spyware or Trojan) on \textsc{Virustotal}~\cite{virustotal}, a famous online antivirus engine.
So far the samples related to our target apps have already been showing up in 20 markets including the leading ones like \texttt{\small Myapp} and \texttt{\small Qihoo 360 Market}.
What's more, \texttt{\small Baidu App Store} keeps receiving apps with this certificate from 2015 to recently -- its latest ``product'' was put on shelf on September 15$^{th}$, 2018.

To this end, we can draw the following conclusions:
(1) Even the leading app markets (and the top developers) are unqualified in detecting malicious apps.
(2) Existing app markets lack information exchange on defending attacks from underground industry.

\begin{table}
	\renewcommand{\arraystretch}{1}
	\small
	\centering
	\caption{Some samples signed by  \protect\\ ``61ed377e85d386a8dfee6b864bd85b0bfaa5af81"}
	\vspace{1mm}
	\rowcolors{2}{gray!25}{white}
	\begin{tabular}{l l c c c c c c}
		\toprule
		{\bf Name} & {\bf Package Name} & {\bf Size} \\
		\midrule
		QQ Talk  & net.in1.smart.qq & 465.8 KB \\
		QQ  & com.h & 8.2 MB \\
		LoveWeChat  & com.lovewechat & 368.4 KB \\
		WeChat  & com.tencen1.mm & 22.1 MB \\
		UC Mini  & com.uc.browser.en & 2.1 MB \\
		UC Browser  & com.UCMobile.microsoft & 21.3 MB \\
		Clean Master  & com.blueflash.kingscleanmaster & 972.0 KB \\
		WiFi Master Key  & com.snda.wifilocating & 5.9 MB \\
		\bottomrule
	\end{tabular}
	\label{table:certificate_case_study}
\end{table}

\begin{table*}
	\renewcommand{\arraystretch}{1}
	\small
	\centering
  \setlength{\belowcaptionskip}{-10pt}
	\caption{Suspicious samples with official certificates}
	\begin{tabular}{l l c c c c c c}
		\toprule
		{\bf Name} & {\bf Sample SHA1} & {\bf Doubtful Point} \\
		\midrule
		iQiyi & b86c55a509e8293b24138b166e9ff410f39e84b5 & Signed by certificate from another developer (360Zhushou) \\
		360Zhushou & 2bb43c53b86d204d0040a8af6cb2a09cf9e93bb7 & Suspicious package name (com.kuyou.sdbgj.baidu) \\
		Youku XL Cracked & b55b7ef189d649aeb03443c5d1ab57c9031d624e & Suspicious word in app name (``Cracked") \\
		\bottomrule
	\end{tabular}
	\label{table:suspicious_samples}
	\vspace{-5mm}
\end{table*}

\noindent{\bf Case study 2.} \emph{Fake samples in different gaming apps}

\begin{figure}
	\centering
	\begin{minipage}[t]{0.165\textwidth}
		\centering
		\includegraphics[width=0.95\textwidth]{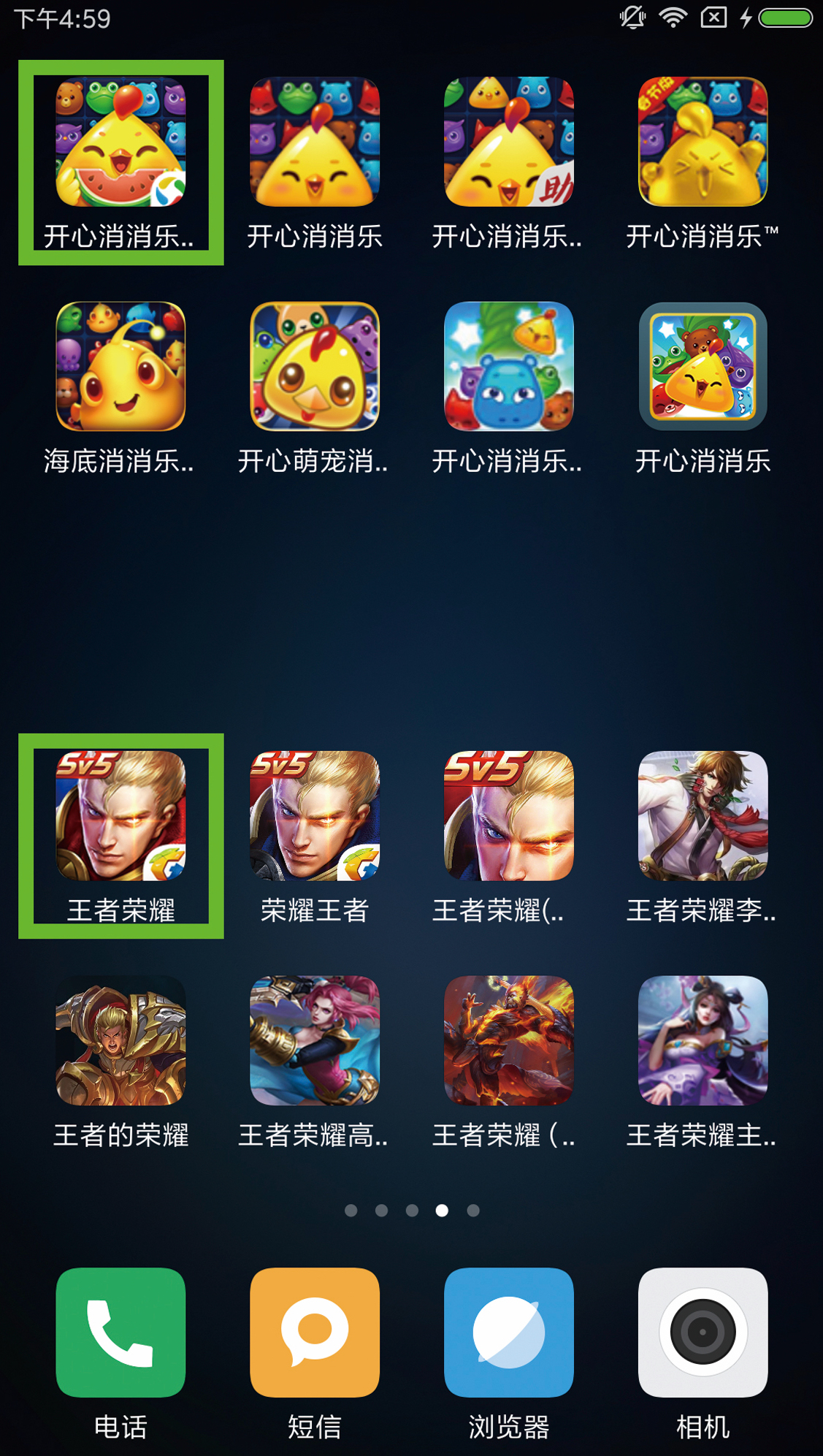}
		\setcaptionwidth{0.9\textwidth}
		\caption{Two games and their fakes}
		\label{fig:screenshot_all}
	\end{minipage}%
	\begin{minipage}[t]{0.165\textwidth}
		\centering
		\includegraphics[width=0.95\textwidth]{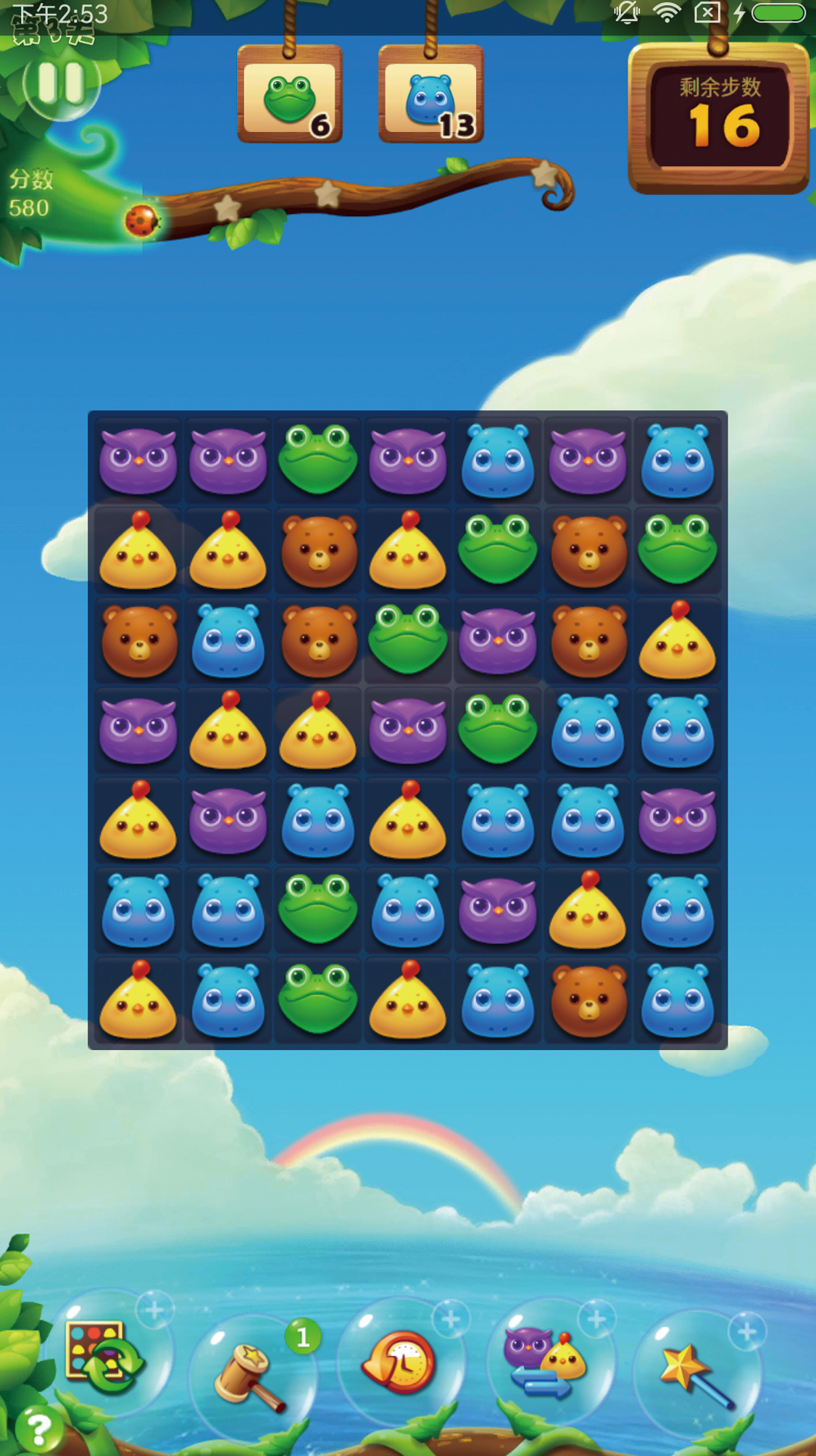}
		\setcaptionwidth{0.9\textwidth}
		\caption{Real \\\textit{\small HappyElements}}
		\label{fig:screenshot_official}
	\end{minipage}%
	\begin{minipage}[t]{0.165\textwidth}
		\centering
		\includegraphics[width=0.95\textwidth]{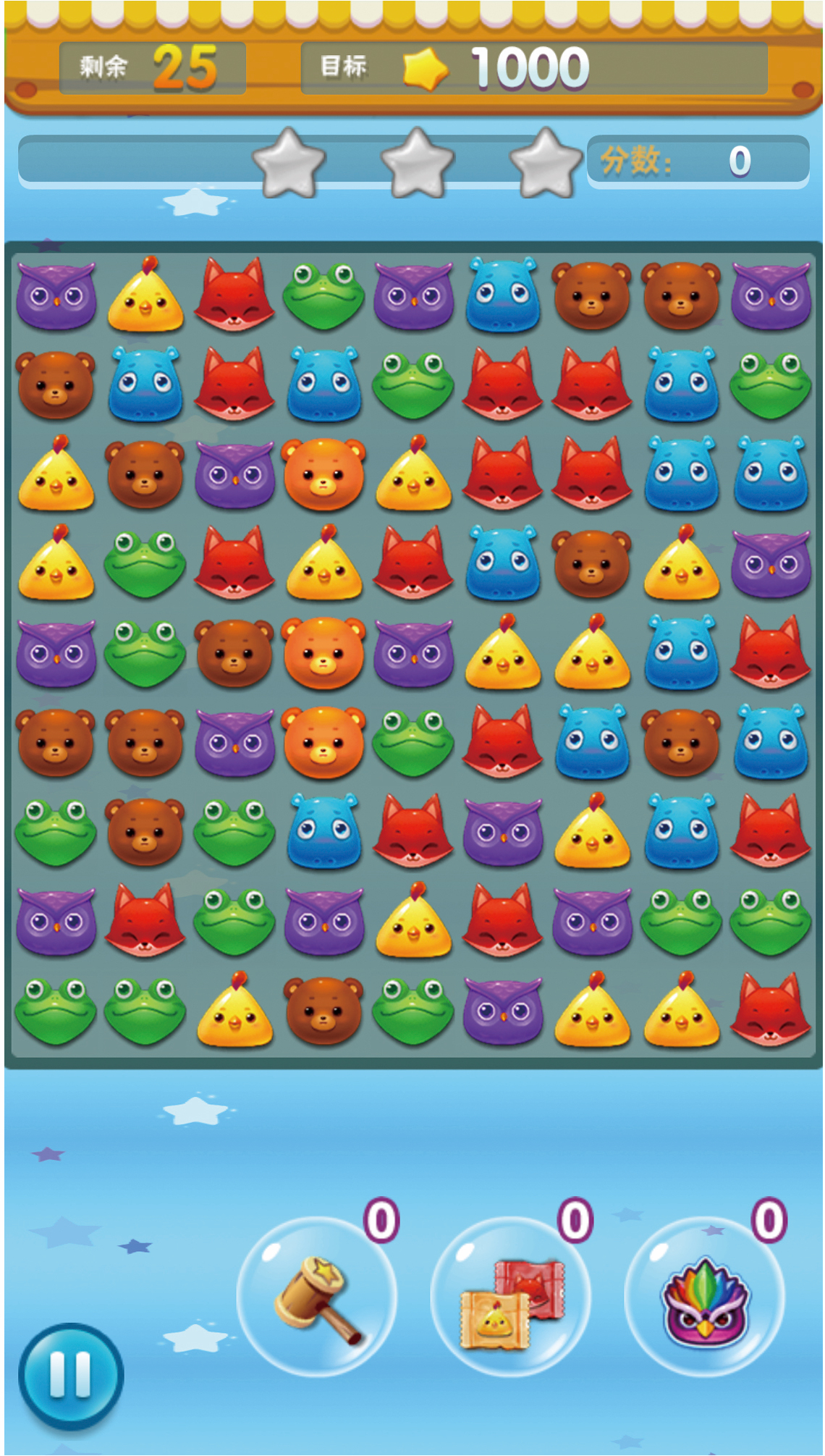}
		\setcaptionwidth{0.9\textwidth}
		\caption{Fake \\\textit{\small HappyElements}}
		\label{fig:screenshot_fake}
	\end{minipage}
\end{figure}

Gaming apps in our target app list (i.e. \texttt{\small ArenaofValor} and \texttt{\small HappyElements}) attract a number of fake samples.
To figure out what do these samples look like, we randomly downloaded some of the fake samples of the 2 gaming apps (7 samples for each) and installed them on our testing device.
Fig.~\ref{fig:screenshot_all} shows how these samples look like on a real Android phone, official apps are marked with green frames.
Apparently, fake samples have either a similar name or a similar icon to official ones.

We even ran these apps on our device.
Screenshots were captured when we ran one of the fake samples (see Fig.~\ref{fig:screenshot_fake}) and the official sample (Fig.~\ref{fig:screenshot_official}).
As a result, we found that 4 fake samples of \texttt{\small HappyElements} are actually games that are similar to the official one (one is a repackaged app with high confidence), 2 are raiders on the game and the last one crashed when it was launched.
3 out of the 4 fake games pop up alert windows in the game to require users for In-App purchase, which is very possible to cause unwilling cost.
All 7 samples are reported to be malicious on \textsc{Virustotal}~\cite{virustotal}.

Fake samples on \texttt{\small ArenaofValor}, in contrast, barely have functionalities like the official one.
3 of those samples are wallpaper setters and the rest 4 are simply puzzle games.
Virustotal reports 6 out of the 7 samples as malware, the last one is claimed as potentially unwanted program (PUP).

We determine it is the difficulty to imitate the official app's functionality that brings about this phenomenon.
The core implementation of multiplayer online battle arena (MOBA) games like \texttt{\small ArenaofValor} is much more complicated than that in \texttt{\small HappyElements}.
What it costs to develop a complex game like \texttt{\small ArenaofValor} is exorbitant for a fake developer.
Therefore, we can infer another reason why apps in \texttt{\small productivity} category gain a low fake sample rate:
Unlike games, on one hand, productivity tools are less likely to have peripheral products (like wallpaper setter mentioned above);
On the other hand, the inevitably laborious developing procedure also prevents the tools themselves from being shammed.

\noindent {\bf Case study 3.} \emph{Suspicious samples with official certificate}

Case study 1 gives us a perfect example on counterintuitive data in our data set.
In order to find out whether or not resemble cases exist in our official samples, we manually reviewed them and noticed a weird entry when sorting out the sample log, a sample claims itself to be ``cracked" in its app name.
Furthermore, we checked (1) if strange word (e.g., ``cracked") appears in our official samples' names, (2) whether or not an official app is signed by an official certificate from another developer, and (3) if one official sample has a suspicious package name.
Eventually we acquired 17 suspicious official samples, listed in table~\ref{table:suspicious_samples} are samples in each of these three kinds .
\textsc{Virustotal} reports that only 2 of the 17 samples are benign, 2 are PUP and the other 13 samples are all malicious.

Despite the possibility that these certificates were somehow leaked to the underground industry, it is more likely that some attackers penetrated the protection scheme.
As far back as December 2017, Google had confirmed and revealed a backdoor on V1 signature scheme (CVE-2017-13156)~\cite{android_security_bulletin}, by which hackers can inject any content into an apk at will without modifying its certificate information.
An alternative solution, V2 signature scheme, has been launched at least one year before that.
In order to confirm if these apps are using the risky V1 scheme, we used a tool, \textsc{apksigner}, provided by Google to verify which signature schemes these samples are using.
It ends up that all 17 samples are using V1 signature scheme.
With actually knowing that V1 is no longer safe, developers still refuse to embrace the safer scheme, which is really disappointing.

\section{Related Work}
\noindent{\bf Empirical study on grayware.}
Andow et al.~\cite{Andow2016ASO} proposed a study of grayware, in which 9 types of greyware are defined and triaged from data retrieved from google play. We referred the definition on \textit{imposter} from this article.

\noindent{\bf Empirical studies on malware ecosystem.}
46 malware samples on various platforms are dissected to gain understanding on their incentive system as in a survey conducted by Felt et al.~\cite{Felt2011ASO}. Meanwhile, several strategies are proposed by them to defend again these type of malware.
Zhou and Jiang~\cite{Zhou2012DissectingAM} gathered over 1,200 malware samples across major Android malware families, systematically characterizing their different natures including installation methods, activation mechanisms and how the payload is carried out.
These researches help expand practitioners' horizon in terms of malicious app's behavior, but the insight they provide may not suit fake app identification well.

\noindent{\bf Repackage detection.}
Prior work on repackage detection generally falls into five categories.
The first one is based on apps' \textit{instruction sequences}, which uses fuzzy hashing techniques to extract the digest of apps, then calculates similarity between every two digests~\cite{DroidMOSS,Zheng2013DroidAnalyticsA}.
The second one is based on \textit{semantic information}.
CLANdroid~\cite{CLANdroid} detects similar apps through analyzing five semantic anchors (e.g., identifiers and Android APIs).
The third kind leverages \textit{lib detection} methods.
CodeMatch~\cite{CodeMatch} filters out libraries used in apps then compares the hash of their remnant.
Wukong~\cite{Wukong} detects repackage apps in two steps, but that it processes the second step by using a counting-based code clone detection approach, instead of hash.
ViewDroid~\cite{ViewDroid} picks out repackage apps by rebuilding and comparing the viewgraph of different apps, belongs to the forth kind which makes use of \textit{visualizes information}.
The fifth kind applies \textit{graph theory} on measuring app similarity.
DNADroid~\cite{DNADroid} calculate apps' similarity based on program dependency graph (PDG), while AnDarwin~\cite{AnDarwin} builds semantic vectors with PDG extracted from every methods.
Centroid~\cite{Centroid} even constructs 3D-control-flow-graph (3D-CFG) for each method in an app and see how alike the centroid in different apps are.

Each of these approaches has its own advantages and drawbacks, from the perspective of scalability and accuracy, which are beyond the topic in this article.
One common they all do share, however, are that the verification step, without any exception, is based on certificate system.
Once the illegal developers poison data with legal certificates through apps with vulnerable signature scheme, even the state of the art detecting approach can do nothing about it.

\vspace{-1mm}
\section{Conclusion}
In this paper we first introduce the concept of fake apps, and study specifically towards these apps.
To the best of our knowledge, we are the first to conduct a comprehensive empirical study on a large-scale fake apps.
To better understand the ecosystem nature of this type of apps, we obtained more than 150,000 data entries from real-world markets, observed and measured the fake samples among this dataset from several dimensions including certificate information, app size, app name and package name, time factor and so on.

Through our measurements we gain valuable experience on fake apps from several perspectives, findings like fake samples' naming tendency and fake developers' evasive strategies are inferred.
To support our findings, we further present a few study cases which provide us a more detailed look into fake apps to back our discoveries on fake app ecosystem.

We hope the lessons learned in this article are informative and helpful for mobile security practitioners in both academia and industry to improve the status quo.
Future work can be processed with the following concerns:
(1) Introducing icon detection into the investigation so that broader samples and higher accuracy can be accessed;
(2) Collecting more data all over the world to examine how global the problem is;
(3) Applying malice detection or even behavior analysis to the samples for qualitative analysis.
As for developers, we advise them to protect their code by obfuscation or encryption, and to use the latest apk signature scheme.
Phantasmagorical and mysterious as the fake apps seem to be, countermeasures will be eventually found as we catch up on their nature.

\vspace{-1mm}
\section*{Acknowledgement}
This research is funded by NSFC Grant 61502170, and the Science and Technology Commission of Shanghai Municipality (No. 18511106202 and No. 18511103802), NTU Research Grant NGF-2017-03-033 and NRF Grant CRDCG2017-S04.

\vspace{-1mm}
\bibliographystyle{IEEEtran}
\balance
\bibliography{seip}

\end{document}